\documentstyle[12pt]{article}
\input amssym.def
\input amssym.tex

\newcommand{\sh}{{\rm sinh}}

\newcommand{\cth}{{\rm coth}}
\newcommand{\be}{{\bf\rm e}}
\newcommand{\bb}{{\bf\rm b}}
\begin{document}

\begin{titlepage}{\LARGE
\begin{center} Why are the rational and hyperbolic\\
Ruijsenaars--Schneider hierarchies\\ 
governed by the same $R$--operators \\
as the Calogero--Moser ones? \end{center}}

\vspace{1.5cm}

\begin{flushleft}{\large Yuri B. SURIS}\end{flushleft} \vspace{1.0cm}
Centre for Complex Systems and Visualization, University of Bremen,\\
Postfach 330 440, 28334 Bremen, Germany\\
e-mail: suris @ mathematik.uni-bremen.de 

\vspace{2.0cm}

{\small {\bf Abstract.} 
We demonstrate that in a certain gauge the Lax matrices of the
rational and hyperbolic Ruijsenaars--Schneider models have a quadratic
$r$-matrix Poisson bracket which is an exact quadratization of the linear
$r$--matrix Poisson bracket of the Calogero--Moser models. 
This phenomenon is explained by a geometric derivation of Lax equations
for arbitrary flows of both hierarchies, which turn out to be governed
by the same dynamical $R$--operator.}
\end{titlepage}

\setcounter{equation}{0}
\section{Introduction}

In the recent years the interest in the Calogero--Moser type of models
\cite{OP}--\cite{R} is considerably revitalized. One of the directions of 
this recent development was connected with the notion of the dynamical 
$r$--matrices and their interpretation in terms of Hamiltonian reduction 
\cite{AT}--\cite{AR}. Very recently, this line of research touched also the 
so--called Ruijsenaars--Schneider models \cite{BB},\cite{AR}, which may be seen 
as relativistic generalizations of the Calogero--Moser ones \cite{RS},\cite{R}. 

In the present paper the same subject as in \cite{AR} is handled, namely, the 
rational and hyperbolic Ruijsenaars--Schneider models. However,
the results are somewhat different, and, as we hope, somewhat more
beautiful. The difference is due to another gauge we choose for the Lax 
matrix of the Ruijsenaars--Schneider models. A striking circumstance comes
out when using our gauge, namely that the both classes of models are governed
by one and the same dynamical $R$--operator. This seems to pass unnoticed in 
the existing literature and is explained in Sect.4. The computations 
presented there are hardly new, at least for the non--relativistic case, but 
we could not find in the literature the main message following from
these computations, namely that they give Lax reprentations for an arbitrary
flow of the corresponding hierarchies, and hence are perfectly suited for
{\it guessing} (not proving!) the correct $r$--matrix ansatz. Sect.3
contains the main result of the paper, namely a quadratic $r$--matrix 
Poisson structure for the rational and hyperbolic Ruijsenaars---Schneider 
models. We compare our results with the previously known ones in the Sect.5.
Besides, for the convenience of a general reader we give a short review of
relevant notions from the $r$--matrix theory in the Sect.2. Sect.6 is devoted
to some problems arising from our results.

\setcounter{equation}{0}
\section{General framework}

In this section we recall some basic notions about integrable hierarchies
and their $r$--matrix theory. Our formulations result from the observations
on the large ''experimental material'' collected in the last decades of
research in this area.

Let ${\cal P}$ be a Poisson manifold; in fact we consider here only the
simplest case of a symplectic space ${\bf R}^{2N}\{x,p\}$ with canonically
conjugated coordinates $x=(x_1,\ldots,x_N)^T$ and $p=(p_1,\ldots,p_N)^T$, so 
that the Poisson brackets of the coordinate functions read:
\begin{equation}\label{can PB}
\{x_k,x_j\}=\{p_k,p_j\}=0,\quad \{x_k,p_j\}=\delta_{kj}.
\end{equation}
Let $H(x,p)$ be a completely integrable Hamiltonian, i.e. suppose that the
hamiltonian system
\begin{equation}\label{Ham syst}
\dot{x}=\{x,H\}, \quad \dot{p}=\{p,H\}
\end{equation}
posesses $N$ functionally independent integrals in involution. Then 
(\ref{Ham syst}) usually (probably always) admits a {\it Lax representation},
i.e. there exist two maps $L:{\cal P}\mapsto g$ and $M:{\cal P}\mapsto g$
into some Lie algebra $g$ such that (\ref{Ham syst}) is equivalent to
\begin{equation}\label{Lax eq}
\dot{L}=[M,L].
\end{equation}
In the cases we are dealing with in this paper, $g=gl(N)$, and it carries
some additional structures. In particular, it is an assosiative algebra with
respect to the usual matrix multiplication, and admits a non--degenerate
scalar product $\langle U,V\rangle={\rm tr}(UV)$. In other cases
$g$ could be a more complicated algebra, for example a loop algebra; then one
would speak about Lax representation with a spectral parameter.

An important observation is that usually
\[
H(x,p)=\varphi(L),
\]
where $\varphi(L)$ is an $Ad$--invariant function on $g$. This is related to
the fact that integarble systems appear not separately, but are organized in
{\it hierarchies}. Namely, to every invariant function $\varphi(L)$ there
corresponds a Lax equation of the form (\ref{Lax eq}). Moreover, there often
holds the following relation:
\begin{equation}\label{M lin}
M=R(\nabla\varphi(L)),
\end{equation}
where $R:g\mapsto g$ is a linear operator, depending, may be, on some of the
coordinates on ${\cal P}$ (it is called then {\it dynamical}). We shall call
$R$ an {\it $R$--operator governing the corresponding hierarchy}. (Recall that 
the gradient $\nabla\varphi(L)$ of a smooth function 
$\varphi$ on an algebra $g$ with a scalar product $\langle U,V\rangle$ is 
defined by the relation
\[
\langle\nabla\varphi(L),U\rangle=
\left.\frac{d}{d\epsilon}\varphi(L+\epsilon U)\right|_{\epsilon=0}, 
\quad\forall U\in g,
\]
and for an $Ad$--invariant $\varphi$ one has $[\nabla\varphi(L),L]=0$). 

Sometimes the relation (\ref{M lin}) has actually another form:
\begin{equation}\label{M quad}
M=R(L\nabla\varphi(L)).
\end{equation}
Of course, in the general setting, where $R$ in (\ref{M lin}) can be dynamic,
the equation (\ref{M quad}) can be seen as $M={\cal R}(\nabla\varphi(L))$ with 
an operator ${\cal R}(\cdot)=R(L\cdot)$. However, if, as it is often the case
in applications, an operator $R$ in (\ref{M quad}) has some special properties
(e.g. is independent on some or all dynamical variables), then it is
advantageous to consider this particular case on its own rights. We shall
call $R$ in (\ref{M quad}) also an {\it $R$--operator governing the 
corresponding hierarchy}, keeping in mind differnce between (\ref{M lin}) and
(\ref{M quad}).

An {\it $r$--matrix theory} \cite{RSTS}, \cite{BV} provides a sort of 
explanation of relations (\ref{M lin}), (\ref{M quad}). Namely, the formula 
(\ref{M lin}) is usually a consequence of a more fundamental fact, namely that 
$L=L(x,p)$ form a Poisson submanifold in $g$ equipped with a so called 
{\it linear $r$--matrix bracket}. This is expressed in the formula
\begin{equation}\label{r Anz}
\{L\stackrel{\otimes}{,}L\}=
\left[I\otimes L,r\right]-\left[L\otimes I,r^*\right].
\end{equation}
The $N^2\times N^2$ matrix $r$ corresponding to the operator $R$ is
defined as 
\[
r=\sum_{i,j,k,m=1}^N r_{ij,km}E_{ij}\otimes E_{km},
\]
where
\begin{equation}
r_{ij,km}=\langle R(E_{ji}),E_{mk}\rangle=
{\rm coeff.\;\;by\;\;}E_{km}{\rm\;\;in\;\;}R(E_{ji}).
\end{equation}
and $r^*$ corresponds in the same way to the operator $R^*$, so that
\[
r^*=\sum_{i,j,k,m=1}^N r_{km,ij}E_{ij}\otimes E_{km}.
\]
In case of constant (non--dynamical) $R$--operators a sufficient condition
for (\ref{r Anz}) to define a Poisson bracket is given by the so called
modified Yang--Baxter equation \cite{RSTS}. In case of dynamical $R$--operators
the corresponding theory is less developed; the most general known sufficient
conditions are given in \cite{Skl}. The lax representation (\ref{Lax eq}) with
$M$ given in (\ref{M lin}) is a consequence of of (\ref{r Anz}) for an
arbitrary $Ad$--invariant function $\varphi(L)$,

Analogously, the formula (\ref{M quad}) is usually a consequence of a more
fundamental fact that $L=L(x,p)$ form a Poisson submanifold in $g$ equipped
with a so called {\it quadratic $r$--matrix bracket}. The most general
quadratic bracket is given by
\begin{equation}\label{as Anz}
\{L\stackrel{\otimes}{,}L\}=(L\otimes L)a_1-a_2(L\otimes L)
+(I\otimes L)s_1(L\otimes I)-(L\otimes I)s_2(I\otimes L).
\end{equation}
Such general quadratic $r$-matrix structures were discovered several times 
independently \cite{FM}, \cite{P}, \cite{S2}; see \cite{S2} for an application
to the Toda and relativistic Toda hierarchy.

In (\ref{as Anz}) the matrices $a_1,a_2,s_1,s_2$ correspond in a canonical way 
to some linear (in principle, dynamical) operators $A_1, A_2, S_1, S_2$ and 
satisfy conditions
\begin{equation}\label{sym}
a_1^*=-a_1,\quad a_2^*=-a_2,\quad s_2^*=s_1,
\end{equation}
and
\begin{equation}\label{sum}
a_1+s_1=a_2+s_2=r.
\end{equation}
The first of these conditions assures the skew--symmetry of the Poisson bracket
(\ref{as Anz}), and the second one garantees that the Hamiltonian flows with
invariant Hamiltonian functions $\varphi(L)$ have the Lax form (\ref{Lax eq})
with the $M$--matrix (\ref{M quad}). If (\ref{sum}) is satisfied, we call the
bracket (\ref{as Anz}) a {\it quadratization} of the bracket (\ref{r Anz}).
In the case of constant operators a sufficient condition for (\ref{as Anz})
with (\ref{sym}), (\ref{sum}) to be a Poisson bracket is validity of the
modified Yang--Baxter equation for three operators $R$, $A_1$, $A_2$; nothing
is known for dynamical case.

It should be remarked that, while in the linear case the correspondence between
operator $R$ in (\ref{M lin}) and matrix $r$ in (\ref{r Anz}) is rather
unambigous, the situation is quite different in the quadratic case. There
exist in principle infinitely many possibilities of choice of $a_1$, $a_2$,
$s_1$, $s_2$ in (\ref{as Anz}), satisfying (\ref{sym}), (\ref{sum}) and
resulting in the same operator $R$ in (\ref{M quad}). All such quadratizations
are parametrized by one skew--symmetric matrix, for example, by $a_1$, because
$a_2=r-r^*-a_1$, $s_1=r-a_1$, $s_2=r^*+a_1$. Hence finding a quadratic
$r$--matrix Poisson structure for a given Lax matrix $L$ is a non--trivial
entertainment even if the $R$--operator governing the whole hierarchy is known.
Guessing a correct quadratization remains more or less a matter of art.
The present paper is devoted to finding such quadratic structure for rational
and hyperbolic Ruisenaars--Schneider models.

\setcounter{equation}{0}
\section{Ruijsenaars--Schneider models  
\newline and their $r$--matrix formulation}

The hyperbolic relativistic Ruijsenaars--Schneider (RS) hierarchy is described 
in terms of the {\it Lax matrix}
\begin{equation}\label{L rel h}
L=L_{\rm RS}(x,p)=\sum_{k,j=1}^N \frac{\sh(\gamma)}{\sh(x_k-x_j+\gamma)}
b_jE_{kj}.
\end{equation}
Here $\gamma$ is a parameter of the model, usually supposed to be pure 
imaginary. We use an abbreviation
\begin{equation}\label{def b h}
b_k=\exp(p_k)
\prod_{j\neq k}\left(1-\frac{\sh^2(\gamma)}{\sh^2(x_k-x_j)}\right)^{1/2},
\end{equation}
so that in the variables $(x,b)$ the canonical Poisson brackets (\ref{can PB}) 
take the form
\begin{equation}\label{rel PB}
\{x_k,x_j\}=0,\quad \{x_k,b_j\}=b_k\delta_{kj}, \quad 
\{b_k,b_j\}=\pi_{kj}b_kb_j
\end{equation}
with
\begin{equation}\label{pi h}
\pi_{kj}=\cth(x_j-x_k+\gamma)-
\cth(x_k-x_j+\gamma)+2(1-\delta_{kj})\cth(x_k-x_j).
\end{equation}

The Hamiltonian of the RS model proper (i.e. of the simplest member of the
hierarchy) is given by
\[
H(x,p)=\sum_{k=1}^N b_k={\rm tr}\;L(x,p).
\]
This model admits a so called non--relativistic limit, achieved
by rescaling $p\mapsto\beta p$, $\gamma\mapsto\beta\gamma$ and 
subsequent sending $\beta\to 0$. In this limit we have $L_{{\rm RS}}=
I+\beta L_{{\rm CM}}+O(\beta^2)$, where the Lax matrix of the rational
Calogero--Moser (CM) hierarchy is introduced:
\begin{equation}\label{L nr h}
L=L_{\rm CM}(x,p)=\sum_{k=1}^Np_kE_{kk}+
\sum_{k\neq j}\frac{\gamma}{\sh(x_k-x_j)}E_{kj}.
\end{equation}
The Hamiltonian of the CM model proper is given by
\[
H(x,p)=\frac{1}{2}\sum_{k=1}^N p_k^2+
\frac{1}{2}\sum_{k\neq j}\frac{\gamma^2}{\sh^2(x_k-x_j)}=
\frac{1}{2}{\rm tr}\;L^2(x,p).
\]

A simple, but important particular case of these models constitute the rational
ones, which are obtained by rescaling $x\mapsto\mu x$, $\gamma\mapsto\mu\gamma$
and sending $\mu\to 0$. In this limit one gets the Lax matrix of the RS 
hierarchy:
\begin{equation}\label{L rel}
L=L_{\rm RS}(x,p)=\sum_{k,j=1}^N \frac{\gamma}{x_k-x_j+\gamma}b_jE_{kj},
\end{equation}
where
\begin{equation}\label{def b}
b_k=\exp(p_k)\prod_{j\neq k}\left(1-\frac{\gamma^2}{(x_k-x_j)^2}\right)^{1/2}.
\end{equation}
The canonical Poisson brackets in terms of $(x,b)$ are still given by 
(\ref{rel PB}) with
\begin{equation}\label{pi}
\pi_{kj}=\frac{1}{x_j-x_k+\gamma}-
\frac{1}{x_k-x_j+\gamma}+\frac{2(1-\delta_{kj})}{x_k-x_j}.
\end{equation}
As a Lax matrix of the CM hierarchy one gets 
\begin{equation}\label{L nr}
L=L_{\rm CM}(x,p)=\sum_{k=1}^Np_kE_{kk}+
\sum_{k\neq j}\frac{\gamma}{x_k-x_j}E_{kj}.
\end{equation}

Now we are in a position to formulate the main result of this paper.

{\bf Theorem.} {\it For the Lax matrices of the relativistic models 
{\rm (\ref{L rel}), (\ref{L rel h})} there holds a quadratic $r$--matrix ansatz 
{\rm (\ref{as Anz})} with the matrices
\[
a_1=a+w,\quad s_1=s-w,
\]
\[
a_2=a+s-s^*-w,\quad s_2=s^*+w,
\]
where in the rational case
\begin{equation}\label{a}
a=\sum_{k\neq j}\frac{1}{x_k-x_j}E_{jk}\otimes E_{kj},
\end{equation}
\begin{equation}\label{s}
s=-\sum_{k\neq j}\frac{1}{x_k-x_j}E_{jk}\otimes E_{kk},
\end{equation}
\begin{equation}\label{w}
w=\sum_{k\neq j}\frac{1}{x_k-x_j}E_{kk}\otimes E_{jj},
\end{equation}
and in the hyperbolic case}
\begin{equation}\label{a h}
a=\sum_{k\neq j}\cth(x_k-x_j)E_{jk}\otimes E_{kj},
\end{equation}
\begin{equation}\label{s h}
s=-\sum_{k\neq j}\frac{1}{\sh(x_k-x_j)}E_{jk}\otimes E_{kk},
\end{equation}
\begin{equation}\label{w h}
w=\sum_{k\neq j}\cth(x_k-x_j)E_{kk}\otimes E_{jj}.
\end{equation}

Let us stress three remarkable properties of the found $r$--matrix structure.
\begin{itemize}
\item All matrices $a$, $s$, $w$ are dynamical, but depend only on the 
coordinates $x_k$, not on the momenta $p_k$.
\item All matrices $a$, $s$, $w$ do not depend on the parameter $\gamma$
of the model.
\item The most important and striking fact is that the structure found is an
exact quadratization of a linear $r$--matrix bracket with
\[
r=a+s,
\]
which is just the $r$--matrix of the non--relativistic CM model found by
Avan--Talon in \cite{AT}. So, the rational and hyperbolic RS hierarchies turn
out to be governed by the same $R$--operators as the rational and hyperbolic
CM hierarchies, respectively. 
\end{itemize}

The proof of the Theorem above is a matter of rather direct computations, and
is therefore omitted. The next section is devoted to a geometric derivation
of the $R$--operators for the RS and CM hierarchies. This derivation is
independent on the our Theorem, and provides some explanation of the third
property mentioned above.

\setcounter{equation}{0}
\section{Derivation of $R$--operator for CM and RS hierarchies}

An important tool to investigate the hyperbolic CM and RS models is, along
with the Lax matrices, an auxiliary diagonal matrix
\begin{equation}\label{X h}
X={\rm diag}(\exp(x_1),\ldots,\exp(x_N))=\sum_{k=1}^N\exp(x_k)E_{kk}.
\end{equation}
The fundamental {\it commutation relation}, describing the structure of the 
Lax matrix $L$ in terms of a given diagonal matrix $X$, reads for the CM model:
\begin{equation}\label{com nr h}
XLX^{-1}-X^{-1}LX=2\gamma\sum_{k\neq j}E_{kj}=2\gamma(\be\be^T-I),
\end{equation}
and for the RS model:
\begin{equation}\label{com rel h}
e^{\gamma}XLX^{-1}-e^{-\gamma}X^{-1}LX=2\sh(\gamma)\be\bb^T.
\end{equation}
Here $I$ stands for the $N\times N$ unity matrix, $\be=(1,\ldots,1)^T$, and
$\bb=(b_1,\ldots,b_N)^T$.

We shall derive the $R$--operators of these models from the results about their
explicit solution obtainable from \cite{OP}, \cite{RS}. These results may be 
formulated as follows. Let $\varphi(L)$ be an $Ad$--invariant function on 
$gl(N)$, and take its value on one of the Lax matrices $L_{\rm CM}$,
$L_{\rm RS}$ as a Hamiltonian function for the corresponding model. 
Then the quantities $\exp(2x_k(t))$ are just the eigenvalues of the matrix
\[
X_0^2\exp(2tf(L_0)),
\]
where 
\begin{equation}\label{f nr}
f(L)=\nabla\varphi(L)\quad{\rm for\;\;the\;\;non-relativistic\;\;CM\;\;case},
\end{equation}
\begin{equation}\label{f rel}
f(L)=L\nabla\varphi(L)=\nabla\varphi(L)L\quad
{\rm for\;\;the\;\;relativistic\;\;RS\;\;case}.
\end{equation}

>From this statement the Lax equations for the corresponding flows can be
derived. Indeed, define the evolution of the matrices $X$, $L$ by the equations
\begin{equation}\label{ev X h}
X^2=X^2(t)=VX_0^2\exp(2tf(L_0))V^{-1},
\end{equation}
\begin{equation}\label{ev L h}
L=L(t)=VL_0V^{-1}.
\end{equation}
Let us explain, in which sence these equations define an evolution. The 
matrix $X^2=X^2(t)$ consists of eigenvalues of the matrix $X_0^2\exp(2tf(L_0))$,
and the matrix $V=V(t)$ is a diagonalizing one. (It is easy to see that the 
matrix $X_0^2\exp(2tf(L_0))$ is similar to a self--adjoint one, if $\gamma$
is pure imaginary). If we fix the ordering of $x_k$ (for example, 
$x_1<\ldots<x_N$), then the only freedom in the definition of $V$ is a left 
multiplication by a diagonal matrix. We fix now $V$ by the condition
\begin{equation}\label{norm V h}
VX_0\be=X\be,
\end{equation}
and show that this assures that the corresponding requirements (\ref{com nr h})
and (\ref{com rel h}) are satisfied for all $t$, provided they were satisfied
for $t=0$. 

Indeed, we have for the CM case:
\[
XLX^{-1}-X^{-1}LX=
X^{-1}VX_0\,\Big(X_0L_0X_0^{-1}-X_0^{-1}L_0X_0\Big)\,X_0^{-1}V^{-1}X
\]
\[
=2\gamma X^{-1}VX_0(\be\be^T-I)X_0^{-1}V^{-1}X=
2\gamma\left(X^{-1}VX_0\be\be^TX_0^{-1}V^{-1}X-I\right).
\]
Since the diagonal entries of the matrix on the left--hand side vanish, we see 
that (\ref{norm V h}) implies $\be^TX_0^{-1}V^{-1}X=\be^T$, which proves the 
commutation relation (\ref{com nr h}) for all $t$. 

Analogously, for the RS case we have:
\[
e^{\gamma}XLX^{-1}-e^{-\gamma}X^{-1}LX
\]
\[
=X^{-1}VX_0\,\Big(e^{\gamma}X_0L_0X_0^{-1}-e^{-\gamma}X_0^{-1}L_0X_0\Big)\,
X_0^{-1}V^{-1}X
\]
\[
=2\sh(\gamma)X^{-1}VX_0\be\bb_0^TX_0^{-1}V^{-1}X.
\]
so that denoting $\bb_0^TX_0^{-1}V^{-1}X=\bb^T$, we see that (\ref{norm V h})
implies the validity of the commutation relation (\ref{com rel h}) for all $t$.

>From this point the derivation of the evolution equations for $L$, $X$
is identical for both the non--relativistic and the relativistic cases. 
Differentiating (\ref{ev L h}), (\ref{ev X h}), we get:
\begin{equation}\label{Lax eq h}
\dot{L}=[M,L],
\end{equation}
(so that the evolution of $L$ is governed by a {\it Lax equation}),
\begin{equation}\label{dot X h}
2X\dot{X}=[M,X^2]+2X^2f(L),
\end{equation}
where
\begin{equation}\label{def M h}
M=\dot{V}V^{-1}.
\end{equation}
In order to find the matrix $M$ explicitly, consider first the off--diagonal
part of (\ref{dot X h}), which implies:
\begin{equation}\label{M kj h}
M_{kj}=\frac{\exp(x_k-x_j)}{\sh(x_k-x_j)}f(L)_{kj}=(1+\cth(x_k-x_j))f(L)_{kj},
\;\;k\neq j.
\end{equation}
The normalizing condition following from (\ref{norm V h}), (\ref{def M h}) reads:
\begin{equation}\label{norm M h}
MX\be=\dot{X}\be,
\end{equation} 
hence
\[
M_{kk}=\dot{x}_k-\sum_{j\neq k}M_{kj}\exp(x_j-x_k).
\]
An expression for $\dot{x}_k$ can be read off the diagonal part of 
(\ref{dot X h}):
\[
\dot{x}_k=f(L)_{kk}.
\]
Substituting this in the previous formula and using (\ref{M kj h}), we 
get:
\begin{equation}\label{M kk h}
M_{kk}=f(L)_{kk}-\sum_{j\neq k}\frac{f(L)_{kj}}{\sh(x_k-x_j)}.
\end{equation}
Finally, notice that we can redefine $M-f(L)$ as a new $M$ (this does
not influence the equations of motion described by the Lax pair), which
results in more convenient expressions
\begin{equation}
M_{kj}=\cth(x_k-x_j)f(L)_{kj},\;\;k\neq j,
\end{equation}
\begin{equation}
M_{kk}=-\sum_{j\neq k}\frac{f(L)_{kj}}{\sh(x_k-x_j)}.
\end{equation}

Analogous results for the rational case may be obtained by the limiting
process, or derived in parallel. An auxiliary diagonal matrix is then given by
\begin{equation}\label{X}
X={\rm diag}(x_1,\ldots,x_N)=\sum_{k=1}^Nx_kE_{kk}.
\end{equation}
Given the diagonal matrix $X$, the structure of the Lax matrix $L$ is completely
described by the following fundamental {\it commutation relations}: for the CM 
case
\begin{equation}\label{com nr}
XL-LX=\gamma\sum_{k\neq j}E_{kj}=\gamma(\be\be^T-I),
\end{equation}
and for the RS case
\begin{equation}\label{com rel}
XL-LX+\gamma L=\gamma \be\bb^T.
\end{equation}

The results about an explicit solution of these models \cite{OP}, \cite{RS} 
read: let $\varphi(L)$ be an $Ad$--invariant function on $gl(N)$, and take its 
value on one of the Lax matrices (\ref{L nr}), (\ref{L rel}) as a Hamiltonian 
function of the corresponding model. 
Then the quantities $x_k(t)$ are just the eigenvalues of the matrix
\[
X_0+tf(L_0),
\]
where (\ref{f nr}), (\ref{f rel}) still hold.
Defining the evolution of the matrices $X$, $L$ by the equations
\begin{equation}\label{ev X}
X=X(t)=V(X_0+tf(L_0))V^{-1},
\end{equation}
\begin{equation}\label{ev L}
L=L(t)=VL_0V^{-1},
\end{equation}
we see that in order to assure that the commutation relations (\ref{com nr})
and (\ref{com rel}) are preserved in the dynamics, one has to normalize
the matrix $V$ by the condition
\begin{equation}\label{norm V}
V\be=\be,
\end{equation}

Now the calculations are again identical for both the CM and the RS cases.
Differentiating (\ref{ev L}), (\ref{ev X}), we get:
\begin{equation}
\dot{L}=[M,L],
\end{equation}
(so that the evolution of $L$ is governed by a {\it Lax equation}),
\begin{equation}\label{dot X}
\dot{X}=[M,X]+f(L),
\end{equation}
where
\begin{equation}\label{def M}
M=\dot{V}V^{-1}.
\end{equation}
Differentiating (\ref{norm V}) gives:
\begin{equation}\label{norm M}
M\be=0,
\end{equation} 

Now we can find the matrix $M$ explicitly. From (\ref{dot X}) we 
immediately obtain the off--diagonal entries of the matrix $M$:
\begin{equation}\label{M kj}
M_{kj}=\frac{f(L)_{kj}}{x_k-x_j},\;\;k\neq j.
\end{equation}
The normalizing condition (\ref{norm M}) implies:
\begin{equation}\label{M kk}
M_{kk}=-\sum_{j\neq k} M_{kj}=-\sum_{j\neq k}\frac{f(L)_{kj}}{x_k-x_j}.
\end{equation}

According to (\ref{f nr}), (\ref{f rel}), we see that the following statement
is proved.

{\bf Proposition.} {\it The general flows of the (rational or hyperbolic) CM 
and RS hierarchies with $Ad$--invariant Hamiltonians $\varphi(L)$ have Lax 
representations (\ref{Lax eq}) with the $M$--matrices given by (\ref{M lin}), 
(\ref{M quad}), respectively. The $R$--operator is {\it one and the same} for 
the CM and RS cases and is given by:
\begin{equation}
R=A+S,
\end{equation}
where $A$ is a skew--symmetric operator on $gl(N)$, and $S$ is a 
non--skew--symmetric one, whose image consists of diagonal matrices.
For the rational models 
\begin{equation}\label{A}
A(E_{kj})=\frac{1-\delta_{kj}}{x_k-x_j}E_{kj},
\end{equation}
\begin{equation}\label{S}
S(E_{kj})=-\frac{1-\delta_{kj}}{x_k-x_j}E_{kk},
\end{equation}
and for the hyperbolic models}
\begin{equation}\label{A h}
A(E_{kj})=(1-\delta_{kj})\cth(x_k-x_j)E_{kj},
\end{equation}
\begin{equation}\label{S h}
S(E_{kj})=-\frac{1-\delta_{kj}}{\sh(x_k-x_j)}E_{kk}.
\end{equation}

Obviously, operators $A$, $S$ canonically correspond to the matrices
$a$, $s$  from (\ref{a}), (\ref{s}) and (\ref{a h}), (\ref{s h}), 
respectively.

\setcounter{equation}{0}
\section{Comparison with the previous results}

As pointed out in the Introduction, the $r$--matrices for the RS models were
previuosly discussed by Babelon--Bernard in \cite{BB} and by Avan--Rollet in
\cite{AR}. In both papers another gauge for the Lax matrix is chosen, namely
a self--adjoint one:
\begin{equation}\label{gauge}
L=\sum_{k,j=1}^N \frac{\sh(\gamma)}{\sh(x_k-x_j+\gamma)}(b_kb_j)^{1/2}E_{kj}.
\end{equation}
Let us reformulate our results for this gauge. Calculations analogous to those 
presented in Sect.4 show that the $M$--matrix for the flow of RS hierarchy 
with a Hamiltonian function $\varphi(L)$ is given in the gauge (\ref{gauge}) by
\[
M_{kj}=\cth(x_k-x_j)f(L)_{kj},\quad j\neq k,
\]
\begin{equation}\label{gauged M}
M_{kk}=-\frac{1}{2}\sum_{j\neq k}\left(\frac{b_j}{b_k}\right)^{1/2}
\frac{f(L)_{kj}-f(L)_{jk}}{\sh(x_k-x_j)},
\end{equation}
where $f(L)=L\nabla\varphi(L)$, as in (\ref{f rel}). Accordingly, the matrices
$a_1$, $a_2$, $s_1$, $s_2$ in a quadratic Poisson bracket (\ref{as Anz}) for
the matrix (\ref{gauge}) depend with necessity not only on $x_k$ but also on
the momenta $p_k$. After direct but tedious calculations one can get a gauged
version of the bracket from our Theorem in the form (\ref{as Anz}) with
\[
a_1=a+\frac{1}{2}(u_1-u_1^*)+v,\quad s_1=\frac{1}{2}(u_2+u_1^*)-v,
\]
\begin{equation}\label{gauged as}
a_2=a+\frac{1}{2}(u_2-u_2^*)-v,\quad s_2=\frac{1}{2}(u_1+u_2^*)+v,
\end{equation}
where
\[
u_1=\sum u_{kj}E_{kj}\otimes E_{kk},\quad u_2=-\sum u_{kj}E_{jk}\otimes E_{kk},
\quad v=\sum v_{kj} E_{kk}\otimes E_{jj},
\]
the coefficients of these matrices being given by
\[
v_{kj}=\frac{1}{4}\Big(
\cth(x_k-x_j+\gamma)-\cth(x_j-x_k+\gamma)+2(1-\delta_{jk})\cth(x_k-x_j)\Big)
\]
and
\[
u_{kj}=\left(\frac{b_j}{b_k}\right)^{1/2}\frac{1-\delta_{jk}}{\sh(x_k-x_j)}.
\]
Let us stress that these matrices, first, depend on momenta $p_k$, second,
depend on the parameter $\gamma$, and third, do not reproduce the $R$--operator
governing the CM hierarchy. This demonsrates how crucially depend the properties
of $r$--matrices on the choice of gauge for Lax matrix.

\subsection{On the Avan--Rollet $r$--matrix}
Avan--Rollet found a {\it linear} $r$--matrix structure of the type 
(\ref{r Anz}) for the Lax matrix (\ref{gauge}). They looked for an $r$--matrix
with entries being linear combinations of elements $L_{kj}$ of the Lax matrix
(\ref{gauge}) with coefficients depending only on coordinates $x_k$, but not on
the momenta $p_k$. According to the remarks above, such linear combinations
can not be cast in the form
\begin{equation}\label{rrr}
r=(L\otimes I)r_1+r_2(L\otimes I),
\end{equation}
necessary for putting linear $r$--matrix structure (\ref{r Anz}) into quadratic
form (\ref{as Anz}). Indeed, in order to have (\ref{rrr}) one is forced to
admit  coefficients depending on momenta.

Actually, after some manipulations based on the explicit expressions of $L_{kj}$,
the Avan--Rollet $r$--matrix can be demonstrated to be equal to (\ref{rrr}) with
\[
r_1=\frac{1}{2}(a+u_1+v),\quad r_2=\frac{1}{2}(a+u_2-v),
\]
which gives back the expressions (\ref{gauged as}). Hence our results and those by
Avan--Rollet are in a sort of {\it ''hidden''} equivalence.

\subsection{On the Babelon--Bernard $r$--matrix}
Babelon--Bernard consider the hyperbolic RS system with a specific value of
parameter $\gamma=i\pi/2$ (which, by the way, precludes the possibility of both
the non--relativistic and the rational reductions). They use the gauge 
(\ref{gauge}) and obtain a quadratic Poisson bracket with $a_1=a_2=\frac{1}{2}a$,
$s_1=s_2=\frac{1}{2}a^{\tau}$, where $\tau$ denotes the transposition in 
the first factor of tensor products. In particular, all $r$--matrix objects
depend only on $x_k$'s. This seems to be in contradiction with the remark after
(\ref{gauged M}). However, a closer look at (\ref{gauged M}) solves this paradox.
For $\gamma=i\pi/2$ the Lax matrix $L$ (\ref{gauge}) is symmetric, which forces
diagonal entries $M_{kk}$ of the matrix $M$ to vanish. It means that for this
particular case the $R$--operator is equal symply to $A$, or, taking into
account the symmetry of $L$, to the $\frac{1}{2}(A+A\circ T)$, where $T$ stands 
for the transposition operator. The structure found by Babelon--Bernard is the
simplest possible quadratization of a {\it twisted} linear $r$--matrix bracket 
with $r=\frac{1}{2}(a+a^{\tau})$. So, it essentially takes care of specific
properties of the model for $\gamma=i\pi/2$ and is not covered by our general
construction.

\setcounter{equation}{0}
\section{Conclusions}

The results of this paper surely constitute only one link in a long chain of 
(already achieved and still hypothetical) results concerning the RS models.
The whole work made for their non--relativistic counterparts should be
repeated, or, better, generalized.
\begin{itemize}
\item After this paper was submitted for publication, two apparently different 
(spectral parameter dependent) quadratic $r$--matrix structures were found for 
the elliptic RS hierarchy \cite{N}, \cite{S}. The structure found in the second
of these preprints serves as a direct generalization of the present results:
it turns out that the elliptic RS hierarchy is also governed by the same 
$R$--operator as the elliptic CM one.
\item The spin Ruijsenaars--Schneider models, introduced recently
in \cite{KZ}, should be also put into the $r$--matrix framework, as it was done 
for the spin Calogero--Moser systems in \cite{BAB}.
\item Another important point is a derivation of our Theorem in the framework 
of Hamiltonian reduction; to this end the results in \cite{G} should be used 
and further developed.
\item It would be rather important to investigate the dynamical analogs of the 
modified Yang--Baxter equation assuring the Poisson bracket properties of the
ansatz (\ref{as Anz}). The investigations of these objects were started in 
\cite{Skl} for the linear ansatz (\ref{r Anz}) and are expected to unveil 
new interesting structures in the case of quadratic brackets. 
\end{itemize}

Further, our Theorem being established, situation with the Calogero--Moser type 
models becomes perfectly analogous to the situation with the Toda--like ones. 
Namely, the transition from the non--relativistic to the relativistic systems 
corresponds to the transition from a linear $r$--matrix Poisson structure to 
its quadratisation. (See \cite{S1}, \cite{S2} for the Toda case). A deeper
understanding of this phenomenon is highly desirable.

\setcounter{equation}{0}
\section{Acknowledgements}

The research of the author is financially supported by the DFG (Deutsche
Forschungsgemeinschaft). 

It is also a pleasure to thank the International
Erwin Schroedinger Institute in Vienna for financial support, 
as well as for the hospitality and nice atmosphere at the 
''Discrete Geometry and Condenced Matter Physics'' workshop, where part of 
this work was done.

Comments by anonymous Referees helped me to improve the presentation.

\end{document}